\documentclass[twocolumn]{aastex63}

\hypersetup{linkcolor=red,citecolor=green,filecolor=cyan,urlcolor=magenta}

\shorttitle{Vortex contribution to the chromospheric heating}
\shortauthors{Yadav et al.}

\graphicspath{{./}{figures/}}

\begin{document}

\title{Simulations Show that Vortex Flows could Heat the Chromosphere in Solar Plage}

\correspondingauthor{Nitin Yadav}
\email{nitnyadv@gmail.com}

\author[0000-0002-8976-516X]{Nitin Yadav}
\affiliation{Max Planck Institute for Solar System Research,
              Justus-von-Liebig-Weg 3, 37077 G\"ottingen, Germany}

\author[0000-0001-9474-8447]{R. H. Cameron}
\affiliation{Max Planck Institute for Solar System Research,
              Justus-von-Liebig-Weg 3, 37077 G\"ottingen, Germany}

\author[0000-0002-3418-8449]{S. K. Solanki}
\affiliation{Max Planck Institute for Solar System Research,
              Justus-von-Liebig-Weg 3, 37077 G\"ottingen, Germany}
\affiliation{School of Space Research, Kyung Hee University, Yongin, Gyeonggi 446-701, Republic of Korea}

\begin{abstract}
The relationship between vortex flows at different spatial scales and their contribution to the energy balance in the chromosphere is not yet fully understood. 
We perform three-dimensional (3D) radiation-magnetohydrodynamic (MHD) simulations of a unipolar solar plage region at a spatial resolution of 10 $\mathrm{km}$ using the MURaM code.
We use the swirling-strength criterion that mainly detects the smallest vortices present in the simulation data.
We additionally degrade our simulation data to smooth-out the smaller vortices, so that also the vortices at larger spatial scales can be detected.
Vortex flows at various spatial scales are found in our simulation data for different effective spatial resolutions. 
We conclude that the observed large vortices are likely clusters of much smaller ones that are not yet resolved by observations.
We show that the vertical Poynting flux decreases rapidly with reduced effective spatial resolutions and is predominantly carried by the horizontal plasma motions rather than vertical flows.
Since the small-scale horizontal motions or the smaller vortices carry most of the energy, the energy transported by vortices deduced from low resolution data is grossly underestimated.
In full resolution simulation data, the Poynting flux contribution due to vortices is more than adequate to compensate for the radiative losses in plage, indicating their importance for chromospheric heating.
\end{abstract}



\section{Introduction} \label{sec:intro}
Vortex flows are ubiquitous and are observed in different layers of the solar atmosphere from the photosphere up to the solar corona.
They have been detected across a broad range of spatial and temporal scales in observations.
Depending on the magnetic environment, their spatial extents, and the layers where they are detected, they are given a variety of names.
A few examples are rotating sun-spots (\citealt{1909evershed,2008yan}), solar tornadoes (\citealt{Wedemeyer-Bohm2012}), giant solar tornadoes (\citealt{Su2012, Wedemeyer2013}), large-scale vortex flows (\citealt{1988brandt,Attie2009,iker2018}) and small-scale vortex flows (\citealt{Bonet2008,bonet2010,ParkS2016}).
There are speculations that they might exist at even smaller scales not yet observed due to observational constraints (\citealt{Giagkiozis_2018,Liu_2019}).

Vortex flows have a significant role to play in both direct current (DC) and alternating current (AC) models of heating of the solar atmosphere.
They twist the footpoints of magnetic flux tubes and thus, excite various MHD waves that propagate to the upper atmospheric layers and eventually dissipate and heat the plasma (\citealt{Fedun2011,Wedemeyer-Bohm2012,kostas_2019}).
Moreover, \citet{Shelyag_2012}, analyzing the photospheric magnetic vortices in simulations, described them as torsional Alfv\'en perturbations propagating along the magnetic field lines.
The estimated Poynting flux contribution of vortex flows suggests that they are a possible channel of transporting the energy needed to heat the chromosphere and the solar corona in quiet Sun conditions (\citealt{Wedemeyer-Bohm2012}).

The vertical Poynting flux provides a measure of electromagnetic energy flux flowing through the solar atmosphere.
The Poynting flux is, on average, positive above the height where $\tau_{\mathrm{Ross}}=1$
(\citealt{steiner_2008,shelyag_2011,abbett_2012}).
The vertical Poynting flux has contributions from both horizontal and vertical plasma motions.
Analysing simulations of a quiet internetwork region, \citet{steiner_2008} proposed that the positive vertical Poynting flux in the photosphere and above is due to the transport of horizontal magnetic fields by vertical flows. 
In contrast, \citet{shelyag_2011} simulated a plage region and suggested that horizontal motions associated with vortex flows are the main source for the vertical Poynting flux. 
\citet{khomenko_2018} performed MHD simulations for both small-scale dynamo and unipolar cases.
They found that in dynamo simulations, the vertical Poynting flux associated with horizontal flows is an order of magnitude larger than the contribution due to vertical flows, whereas, in their unipolar simulations they found similar contributions from both horizontal and the vertical flows.
Thus, the contribution of the horizontal and the vertical flows largely depend on the magnetic configuration of the solar atmosphere.

In addition to observations, vortex flows have also been widely detected in numerical simulations (\citealt{Nordlund1985,moll2012,Shelyag_2012,Kitiashvili_2012}).
However, most of the simulation studies pertain to the quiet Sun conditions.
Additionally, vortex detection in simulations is strongly influenced by the spatial resolution of simulation data and vortex detection methods.
Almost all the existing vortex detection methods are based on the gradients of velocity fields in neighboring pixels.
Therefore, they detect strongly rotating but smaller vortices and often miss the more slowly rotating large scale vortices.
Therefore, in the aforementioned simulations, mainly the smallest vortices present in the data were detected and the larger vortex flows (those found in the observations) remain undetected.

Recent statistical studies suggest that vortices have continuous distributions of lifetimes and sizes, with short-lived smaller vortices being the most abundant (\citealt{Giagkiozis_2018,Liu_2019,Silva_2018}).
These authors have applied the local correlation tracking (LCT) technique to observed intensity maps, which might not detect smaller, short-lived vortices.
Therefore, the estimated contribution of vortices to the total energy flux will be inaccurate.
To quantify the role of vortex flows in the energy transport, it is important to include the contribution of vortex flows of all scales, also from the vortices that are smaller than the spatial resolution of currently available observational instrumentation.
One way to overcome the limitation of current observations is to look for a scaling law for the energy contribution of the vortices with respect to their spatial or temporal scales using comprehensive numerical simulations. Then their total energy contribution can be estimated even at unresolved scales.
\begin{figure}
    \centering
    \includegraphics[scale=0.45,trim=0.5cm 0 0 0]{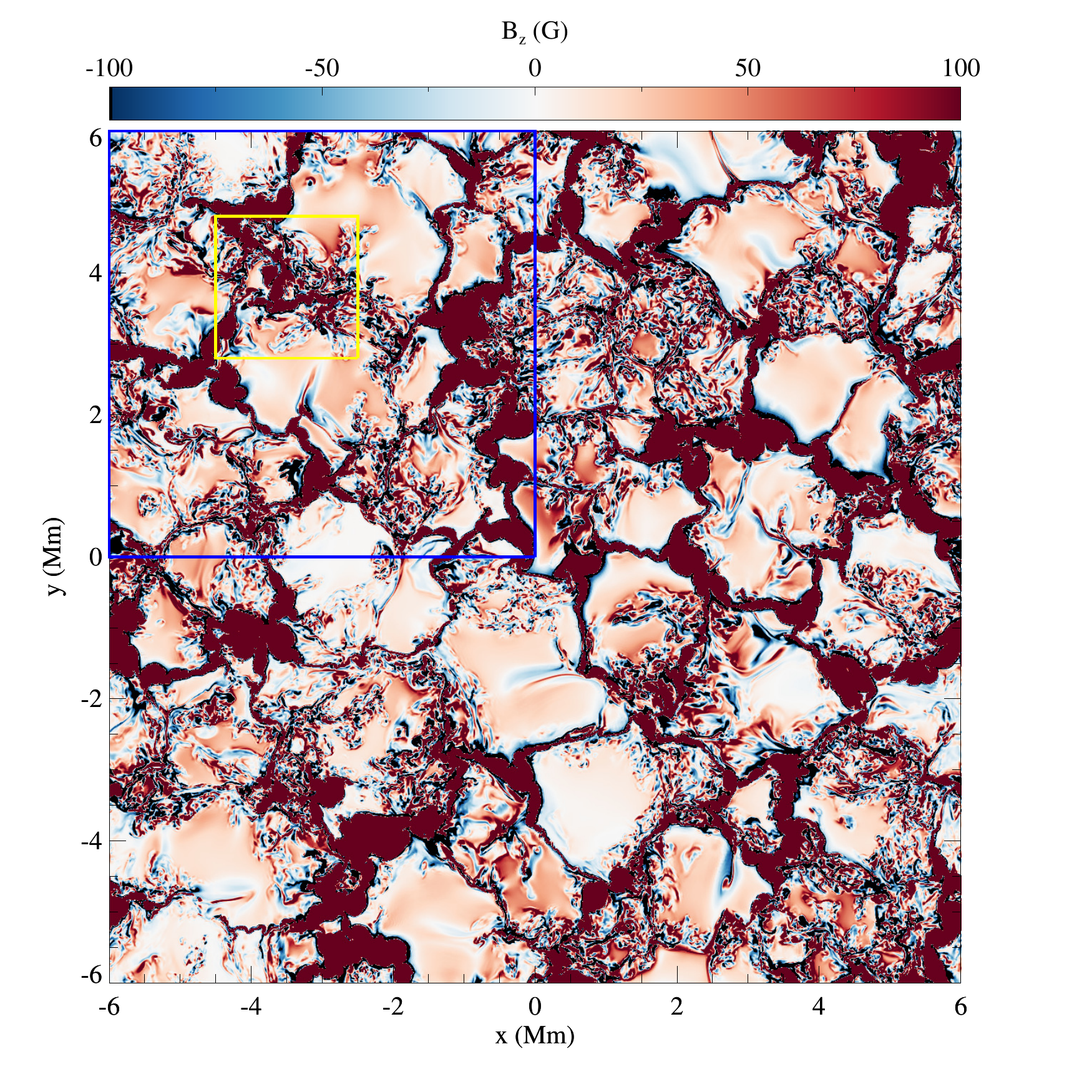}
    \caption{Spatial map of the vertical component of the magnetic field strength at the continuum formation height (saturated at $\pm$100 G). The blue and yellow square  encloses the regions displayed in the left and right column, respectively, in Fig. \ref{fig:my_label2}.}
    \label{fig:my_label1}
\end{figure}
\begin{figure*}
    \centering
    \includegraphics[scale=0.55]{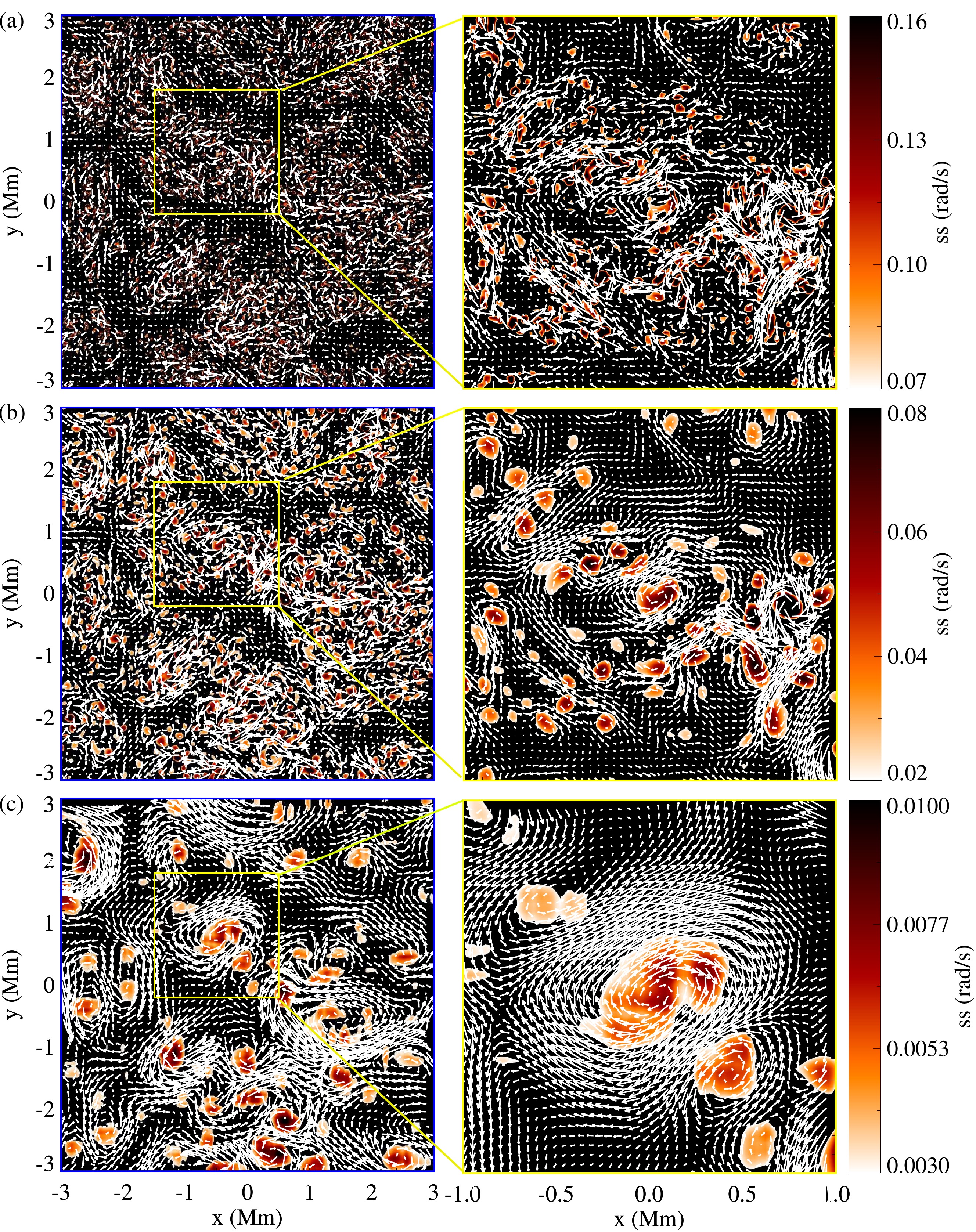}
    \caption{Horizontal velocity vector at a height of 1 $\mathrm{Mm}$ above the mean solar surface  at various effective spatial resolutions viz. 10 $\mathrm{km}$ (a), 100 $\mathrm{km}$ (b) and 500 $\mathrm{km}$ (c). The plotted area corresponds to the region lying in the blue square in Fig. \ref{fig:my_label1} (left column) and zoomed to cover the region marked by the yellow square (right column). The color signifies the swirling strength (ss), see the colour bars.}
    \label{fig:my_label2}
\end{figure*}

In this letter, we aim, with the help of numerical simulations, at answering the following questions: Are there vortex flows present at multiple spatial scales? If yes, how are they related with the spatial resolution of the simulations? Is there any scaling law existing for the vertical Poynting flux with spatial resolution? Which plasma flows are the main contributor for the vertical Poynting flux in the plage chromosphere? What is the quantitative contribution of vortices to the total energy transport through various layers of the solar atmosphere in a plage region?
Is it sufficient to heat the chromosphere in solar plage? 
After this introduction, we describe the numerical simulation details in Sect. \ref{2}. 
Results are discussed in Sect. \ref{3}, followed by a summary and conclusions in Sect. \ref{4}.
\section{Methodology}\label{2}
We have performed 3D radiation-MHD simulations of a unipolar solar plage region using the MURaM code (\citealt{voegler2005,Rempel2014,Rempel2017}). 
The code solves the  coupled  system  of the compressible MHD equations including non-grey radiative transfer.
Our simulations do not include non-LTE effects, however, they incorporate the effects of density stratification, and realistic convective driving at the sub-surface layers by turbulent convection.
Since the waves are generated by the photospheric turbulence, investigating their properties in these simulations may give us guidance on their role in the real Sun.
The simulation box has dimensions of 12 $\mathrm{Mm}$ $\times$ 12 $\mathrm{Mm}$ $\times$ 4 $\mathrm{Mm}$ (x, y, z, where z is normal to the solar surface, pointing outward).
The mean solar surface (continuum formation height) is located 1.5 $\mathrm{Mm}$ above the bottom of the simulation box. The spatial grid spacing is 10 $\mathrm{km}$ in all the three directions.
Periodic boundary conditions are assumed for the lateral boundaries.
The magnetic field is vertical at the top and bottom boundaries.
For plasma flows, the top boundary is closed and allows horizontal motions.
The bottom boundary allows both in- and outflows.
The top boundary is located at a height of 2.5 $\mathrm{Mm}$ above the mean solar surface.
It is closed to vertical flows and partly reflects the motions that reach the top of the simulation box.
This is not so unreasonable as plage is generally found at one footpoint of a coronal loop and a downward flux can be expected from the other end of the loop as it reenters the photosphere.
In a first step, the simulation is run without a magnetic field (purely hydrodynamic case) for almost 2 hrs, by when it has reached a statistically stationary state.
Then we introduce a vertical magnetic field of 200 G to simulate a plage region and ran it again for 1.2 hrs of solar time.
After that we saved data from a 5 minute sequence at 10 s cadence and used that for the further analysis.
In recent years, it has been shown that ambipolar diffusion may be of significance for chromospheric heating (\citealt{khomenko_2012,Shelyag_2016,khomenko_2018}).
\citet{khomenko_2018} found that Poynting flux absorption by ambipolar diffusion depends strongly on the complexity of magnetic field.
They showed that ambipolar diffusion is more effective in quiet sun regions where mixed polarity fields are present than in more regular and unipolar regions.
Therefore, in our current study of plage regions, with relatively ordered field, we neglect ambipolar diffusion.

To identify vortices we apply the swirling-strength criterion (\citealt{zhou1999}) that provides the angular velocity of rotating plasma.
The swirling strength criterion has been shown as superior to standard method of enhanced vorticity (\citealt{Kato2017}).
Since this method as well as other vortex identification methods use velocity gradients to identify vortex locations, they extract the smallest vortices present in the simulation data.
To identify the larger vortices present in the simulation data, we degrade our simulation data so that the smaller vortices are smeared out, whereas, the larger vortices survive. 
To do this, we convolve the simulation data with Gaussian kernels having a FWHM of 50 $\mathrm{km}$, 100 $\mathrm{km}$, 250 $\mathrm{km}$ and 500 $\mathrm{km}$, respectively and then apply the swirling-strength criterion on the resulting degraded data sets to detect the larger vortices.
\section{Results}\label{3}
Figure \ref{fig:my_label1} displays a spatial map (saturated between $\pm 100$ G ) of the vertical component of the magnetic field at the continuum formation height.
The magnetic field strength in strong magnetic elements reaches up to kiloGauss values (dark red patches) and small-scale, weaker mixed polarity regions are formed in between due to the turbulent convection.
To compare the vortex flows at various spatial scales, we select a horizontal slice of 6 $\mathrm{Mm}$ $\times$ 6 $\mathrm{Mm}$ at a height of 1 $\mathrm{Mm}$ above the mean solar surface (corresponding to the blue square in Fig. \ref{fig:my_label1}) and display the horizontal component of the velocity vectors at full resolution as well as for the degraded simulations in the left column of Fig. \ref{fig:my_label2}.
Interestingly, at full resolution (panel a) we do not see any large-scale rotational motion in the chromosphere.
The larger scale vortices only become evident after degrading the data to lower resolution (panel c).
This is because the simulation data at high-resolution are dominated by numerous small-scale structures which makes the presence of large scale structures indiscernible.
Vortices at all scales smaller than the simulation box are possibly formed by a local imbalance in the vorticity (which has an overall zero mean in these simulations) driven by turbulence.
The small-scale vortices inside a large-scale vortex are often of opposite sign, with only the net vorticity appearing at the larger scales.
\begin{figure}
    \centering
    \includegraphics[scale=0.48,trim=0 0 0 0]{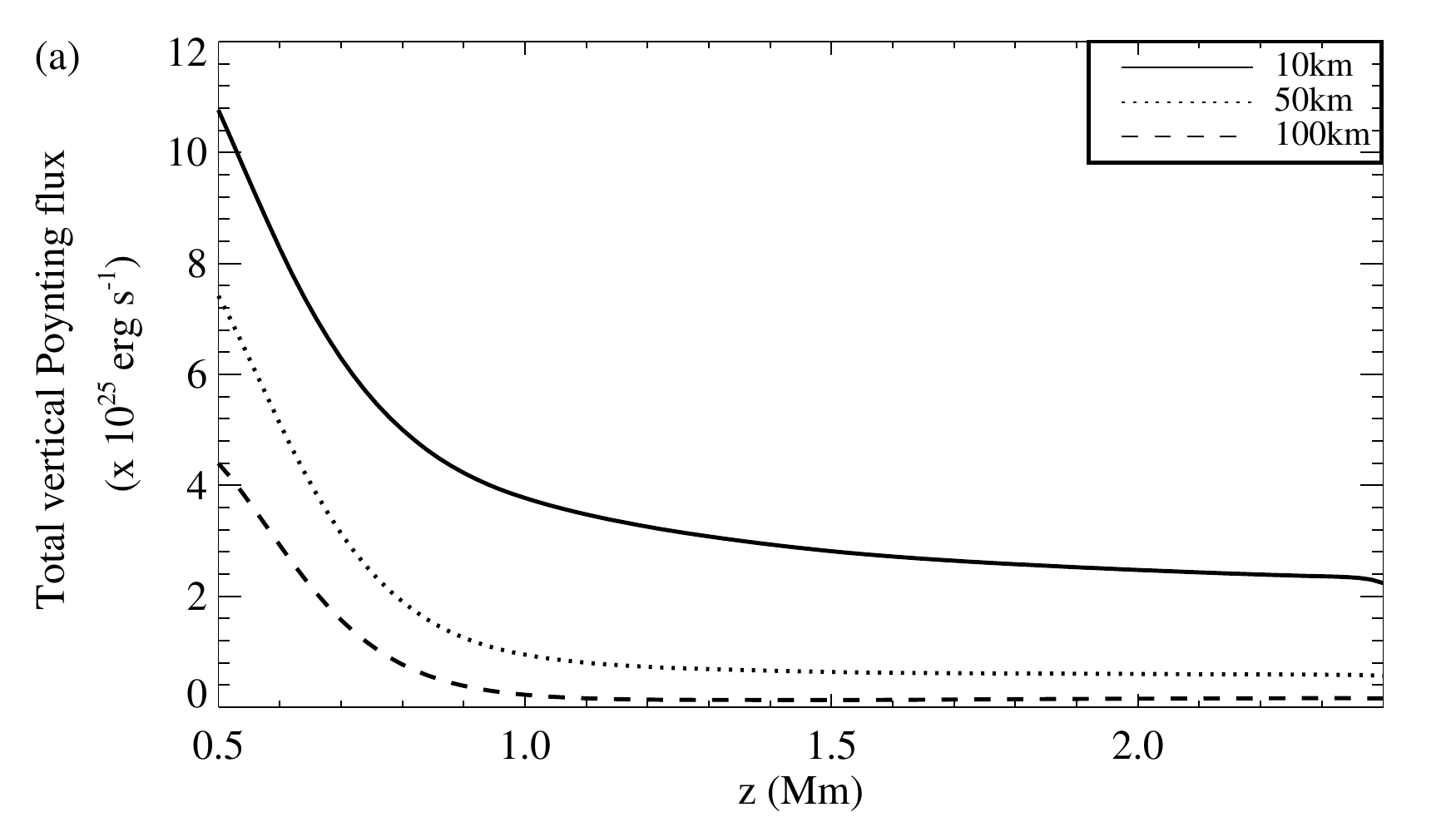}
    \includegraphics[scale=0.48,trim=0 0 0 0]{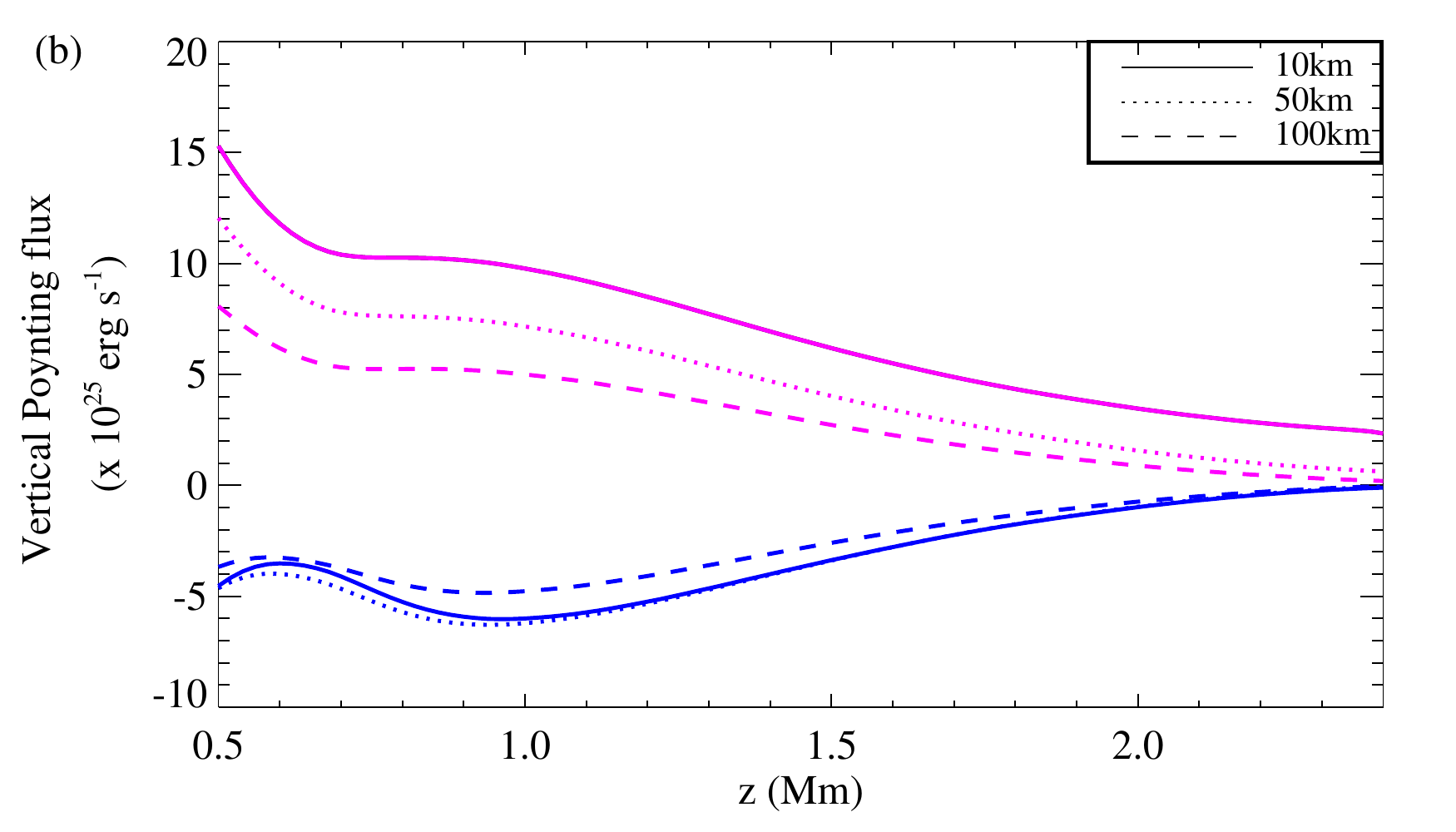}
    \includegraphics[scale=0.48,trim=0 0 0 0]{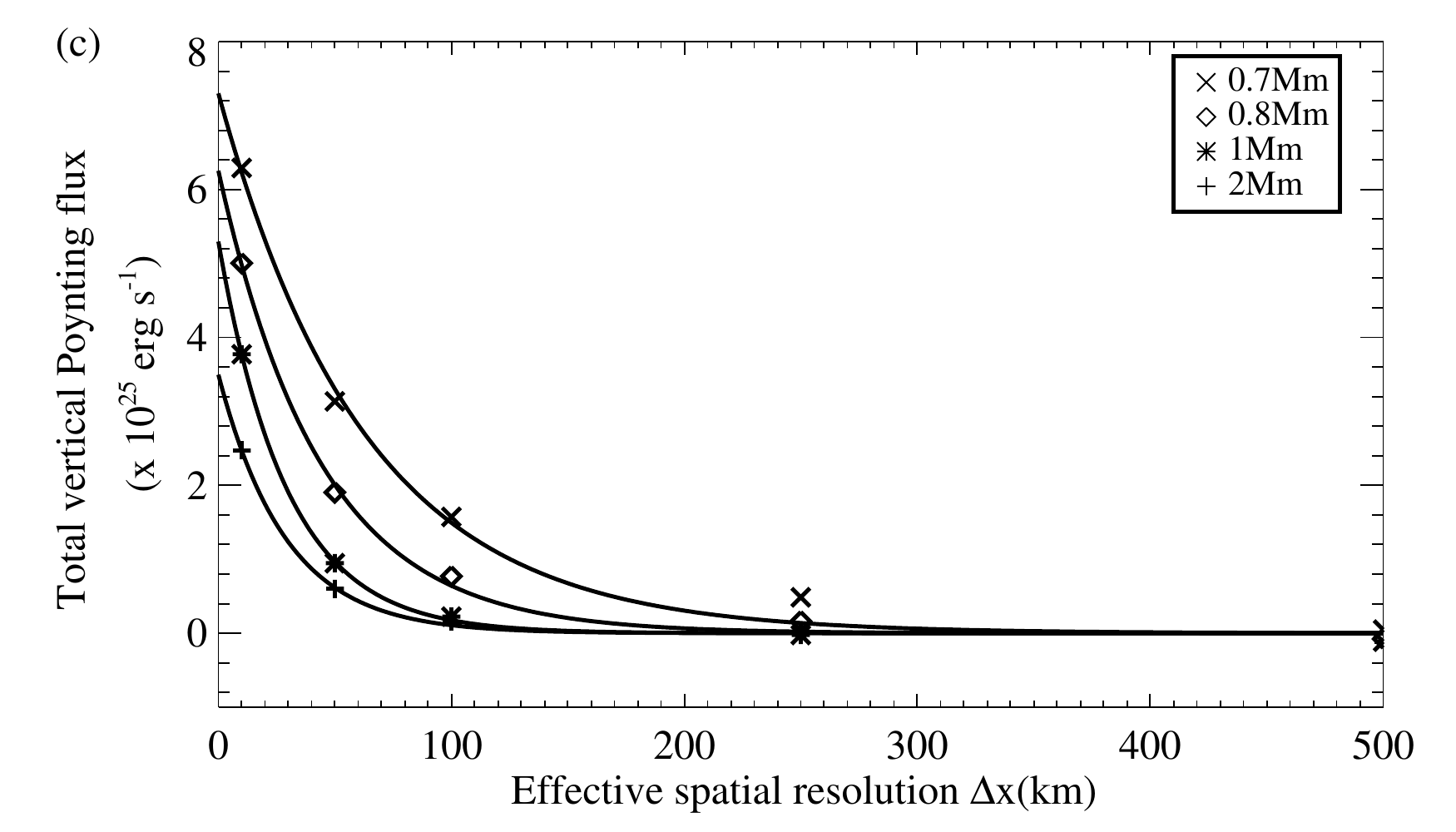}
\caption{(a) Total net vertical Poynting flux in the chromosphere for different resolutions, (b) Part of total vertical Poynting flux generated due to horizontal (magenta) and vertical (blue) plasma motions, (c) Total net vertical Poynting flux at various heights above the mean solar surface at various resolutions (see legend for the exact heights the various symbols correspond to) and exponential fit as a function of effective spatial resolution (solid line).}
    \label{fig:my_label3}
\end{figure}
\begin{figure}
    \centering
    \includegraphics[scale=0.25]{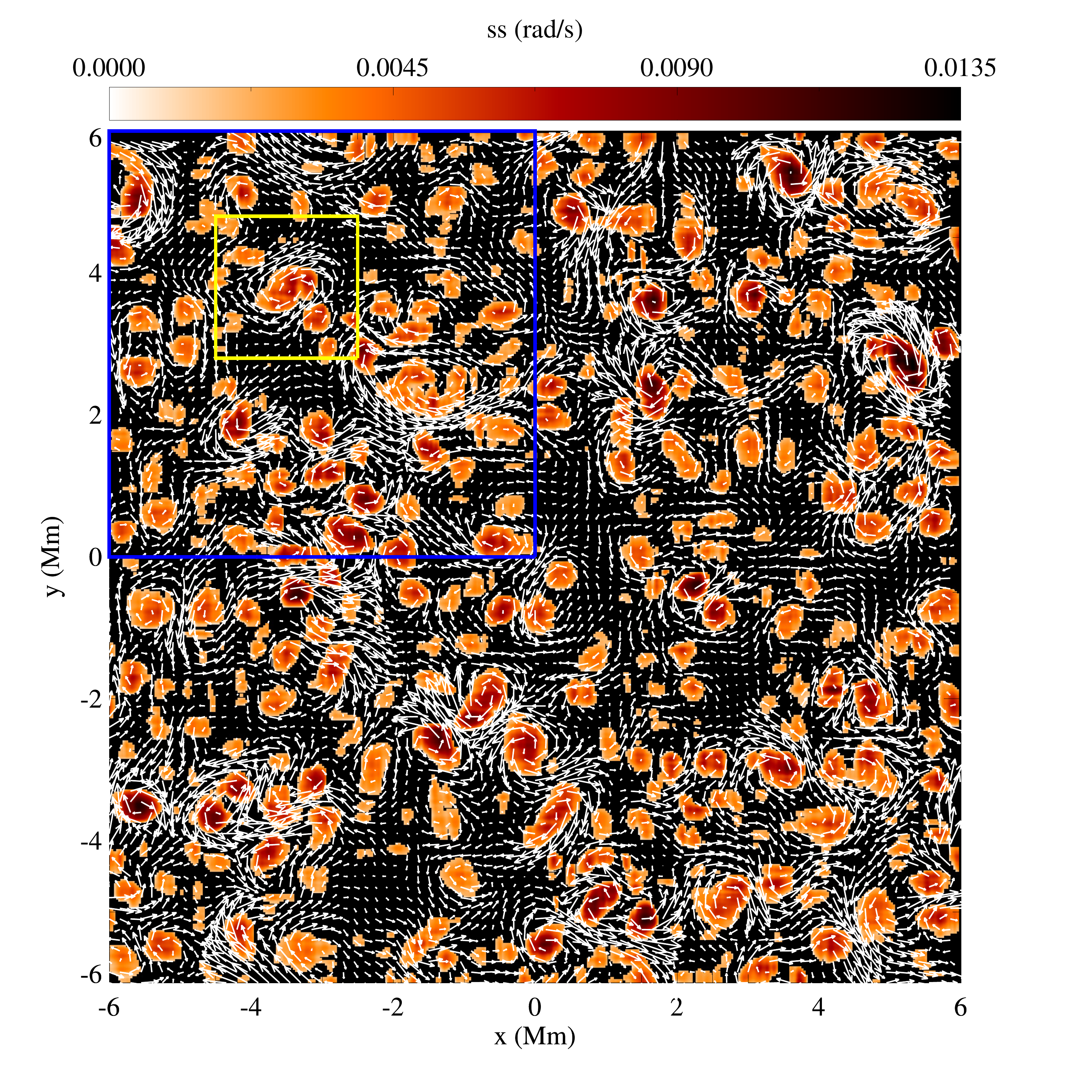}
    \caption{Horizontal velocity vector at a height of 1 $\mathrm{Mm}$ above the mean solar surface for the degraded data with a spatial resolution of 500 $\mathrm{km}$. Color represents swirling strength for the contiguous vortices selected by the swirling strength criterion for further analysis. Blue and yellow squares indicate the regions displayed in the left and right column, respectively, in Fig. \ref{fig:my_label2}.}
    \label{fig:my_label5}
\end{figure}
\begin{figure}
    \centering
    \includegraphics[scale=0.4,trim=0 1.7cm 0 0]{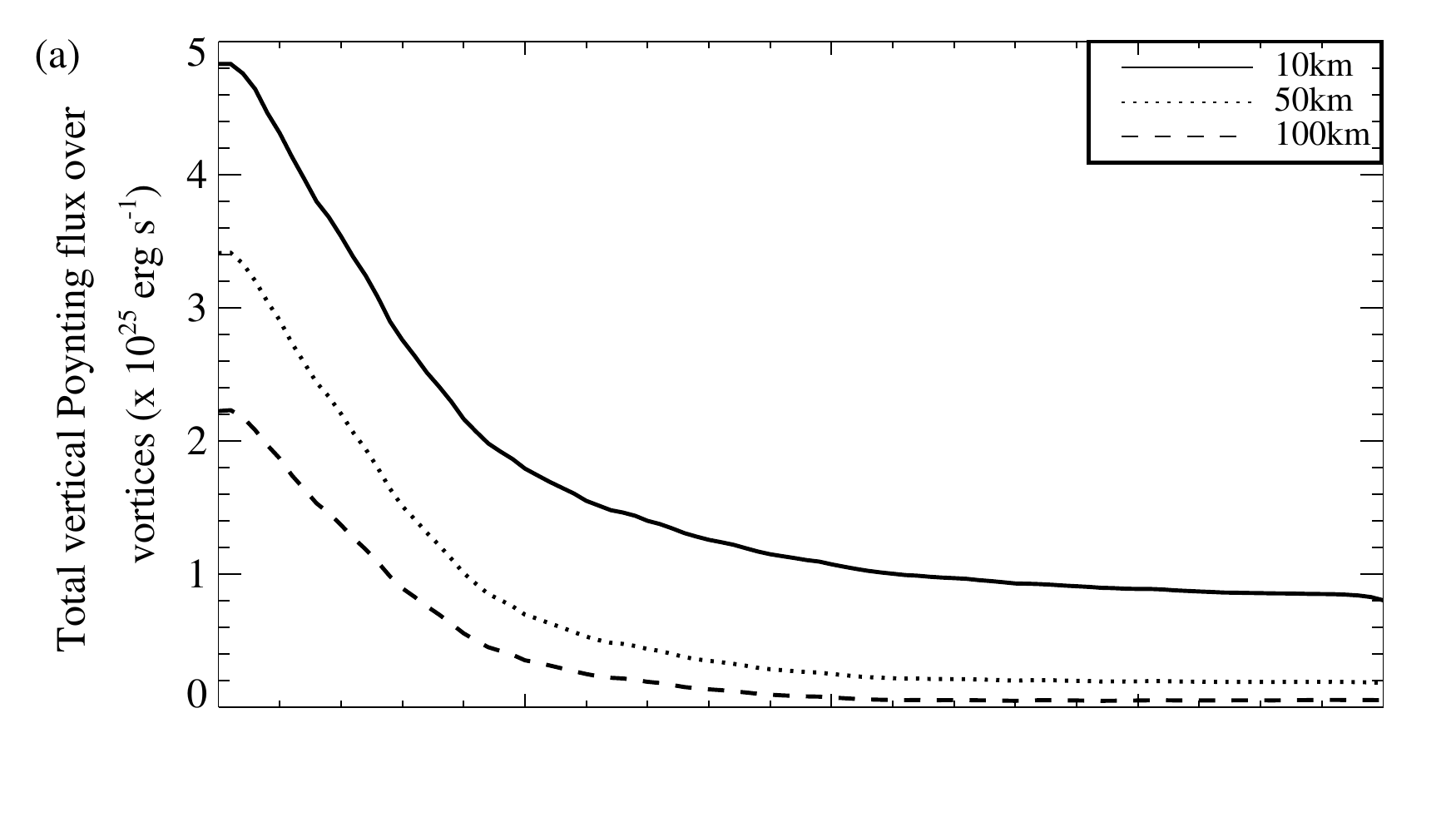}
    \includegraphics[scale=0.4,trim=0 1.7cm 0 0]{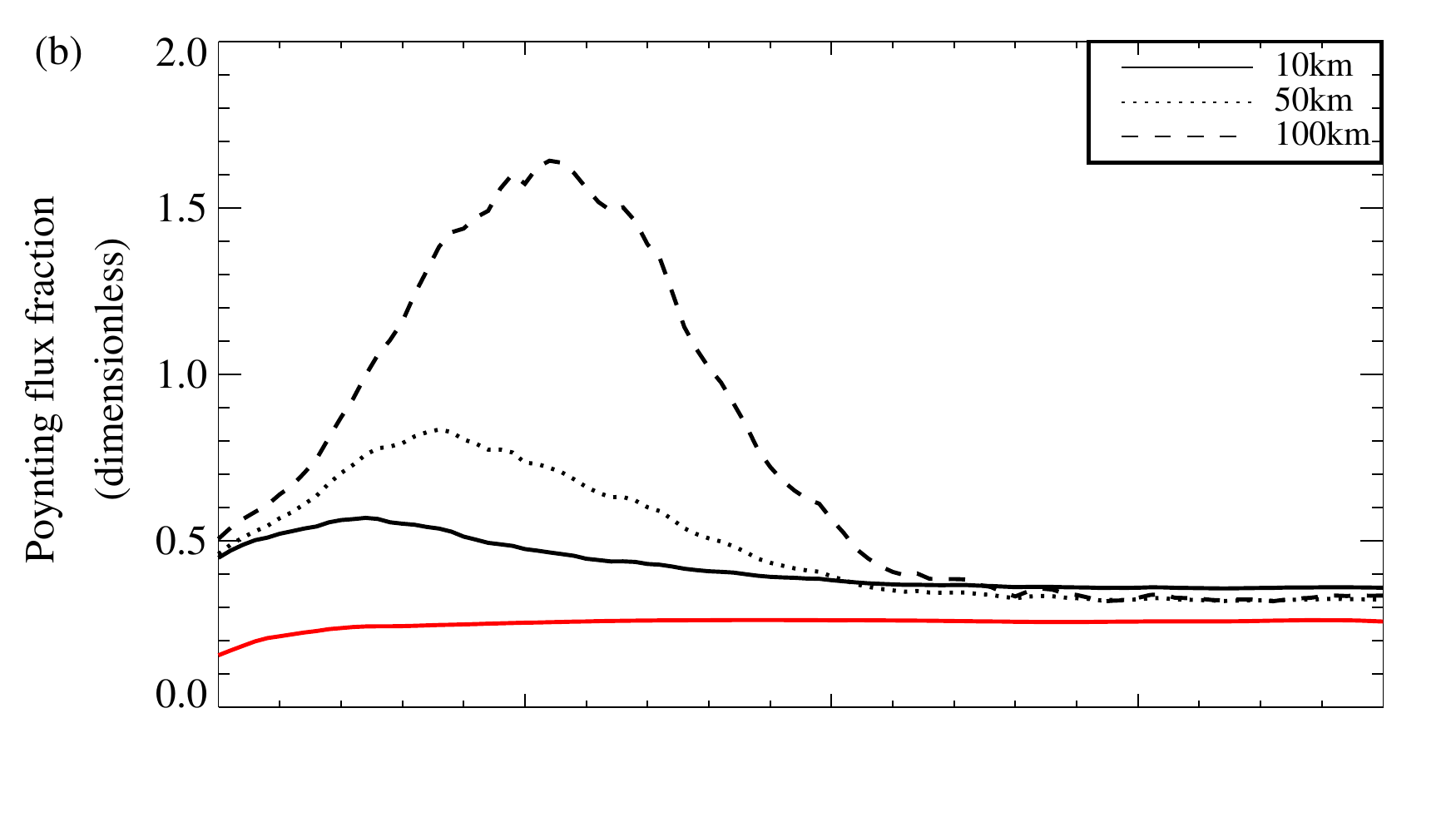}
    \includegraphics[scale=0.4,trim=0 0 0 0]{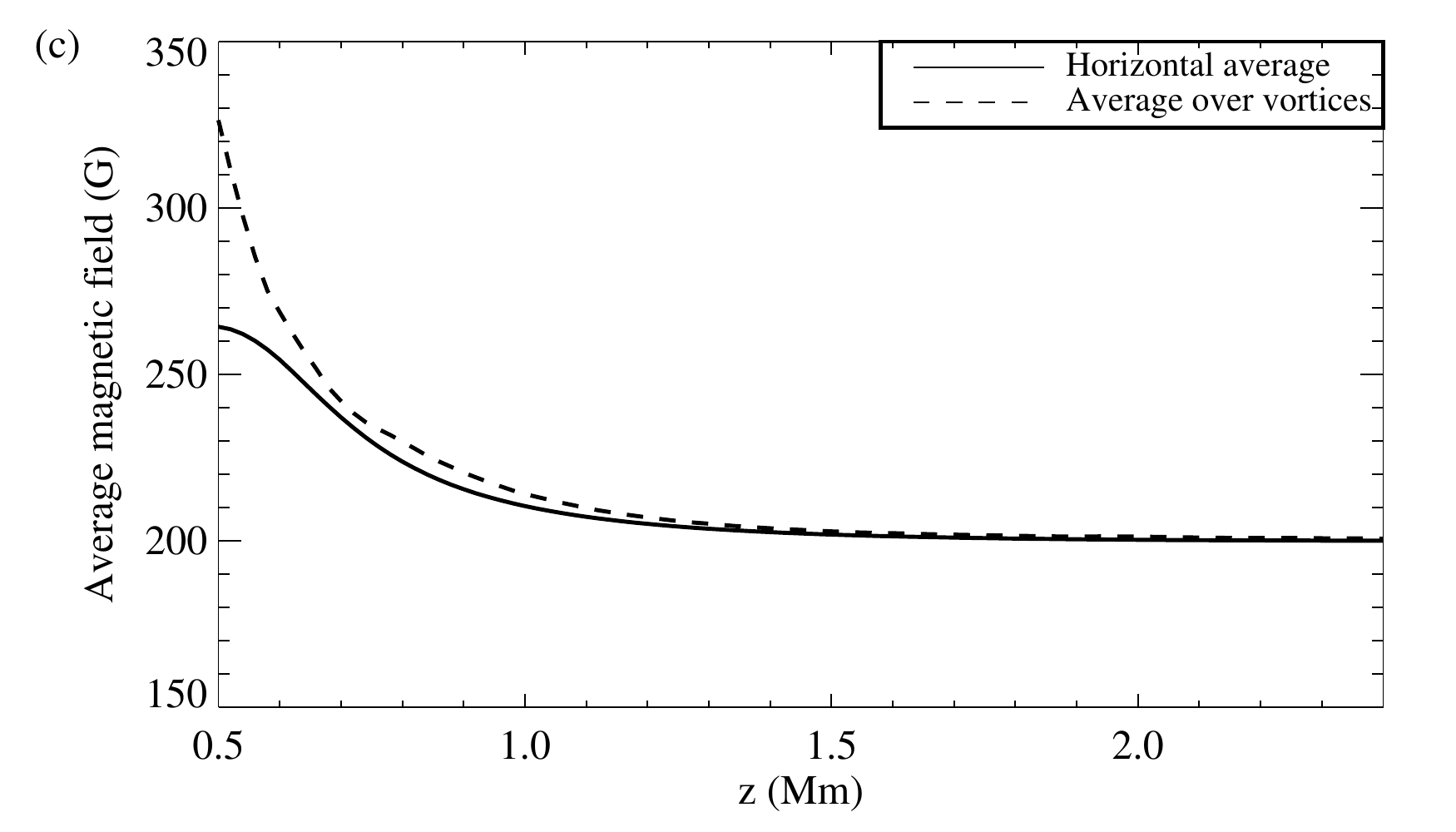}
    \caption{(a) Total net vertical Poynting flux over vortex locations for different resolutions, (b) Relative Poynting flux over vortex locations (normalized by the total Poynting flux at the same geometrical height) and area fraction covered by vortices (red), (c) Averaged magnetic field strength over horizontal domain (solid curve) and over vortex locations (dashed curve) at full resolution simulation data.}
    \label{fig:my_label6}
\end{figure}

To examine the sizes of these rotating plasma structures at different effective spatial resolutions, we consider, as an example, a single vortex seen at 500 $\mathrm{km}$ resolution (lying at the center of the yellow square) in the bottom left panel. The panels on the right show blow-ups of this yellow square, at the three selected resolutions. 
Note that at full resolution of the simulation, there are numerous vortices identified by the swirling strength criterion in the yellow box (top right panel). They have a horizontal extension of 100-200 $\mathrm{km}$.
In the degraded simulation data with 100 $\mathrm{km}$ effective resolution, we find larger rotating structures with a typical width of around 200-400 $\mathrm{km}$, and in the degraded data with 500 $\mathrm{km}$ effective resolution, we find vortex flows having a diameter of $\sim$ 1.5 $\mathrm{Mm}$, which is comparable to the observed chromospheric swirls (\citealt{Wedemeyer-Bohm2012}). 
Consequently, we argue that these large observed vortices are actually composed of many small-scale vortices that have not yet been observed due to limited spatial resolution of observations.
Also, the swirling strength of smaller vortices is much higher than the large scale vortices, implying that they rotate much faster than their larger counterparts.

In our further analysis we perform a statistical investigation by taking the temporal mean over 30 snapshots spanning a 5 minute sequence.
We calculate the total vertical Poynting flux as well as its components due to the horizontal and the vertical plasma motions for data sets at various resolutions.
Panel (a) of Fig. \ref{fig:my_label3} displays the variation of horizontally averaged vertical Poynting flux with height.
With improved resolution, the Poynting flux increases because of the contribution from small scale motions that are averaged out in the degradation process.
Panel (b) of Fig. \ref{fig:my_label3} compares the contribution of horizontal (magenta) and vertical (blue) plasma motions to the total net vertical Poynting flux.
We see that Poynting flux is predominantly carried by the horizontal plasma motions which essentially underlines the importance of vortex flows in the energy transport as vortices are generally associated with large horizontal plasma flows.

It it evident from Fig. \ref{fig:my_label3}(a) that the total vertical Poynting flux is strongly influenced by the effective spatial resolution.
Therefore, it is important to explore if there is a scaling law, so that results can be extrapolated to as yet unresolved scales.
Using the total vertical Poynting flux at various resolutions at any given height, we find a least squares best-fitting curve of the form $S_z=S_oe^{-a\Delta x}$, where  $S_0$ and $a$ are the amplitude and the exponent (fit parameters), respectively, and $\Delta x$ is the effective spatial resolution.
We display the total vertical Poynting flux for various effective spatial resolutions (symbols) at heights of 700 $\mathrm{km}$, 800 $\mathrm{km}$, 1 $\mathrm{Mm}$ and 2 $\mathrm{Mm}$ above the mean solar surface in Fig. \ref{fig:my_label3}(c).
The best fit exponential curve (solid line) is also overplotted.
The amplitudes are 7.3, 6.2, 5.3, and 3.5, respectively, and the exponents are -0.015, -0.022, -0.03, and -0.03, respectively, for each of these heights.
Assuming that the fit extends to higher resolutions (i.e. $\Delta x$ $\rightarrow$ 0) then Fig. \ref{fig:my_label3}(c) indicates that a spatial resolution of 10 $\mathrm{km}$, as used in the present case, is sufficiently high for energy estimates.

Next, we investigate the explicit contribution of vortices to the energy transport.
As an illustrative example, we use the degraded data set with an effective spatial resolution of 500 $\mathrm{km}$.
Using the regions covered by larger vortices is advantageous as that will provide the accumulative contribution of the smaller and the larger vortices (that are present in the simulation data sets with better resolutions but remain undetected due to their comparatively weaker swirling strength than smaller vortices) to the total energy transport.
To select the vortex regions we apply swirling strength threshold of 0.002 rad s$^{-1}$ that correspond to the regions that have swirling period less than 50 minutes.
This threshold is chosen such that 
the selected regions encompass most of the rotating structures visually identified by horizontal velocity vectors. 
We also require that each 3D vortex structure consists of at least 100 grid points so as to avoid smaller transient structures.
Figure \ref{fig:my_label5} displays the swirling strength at the selected regions by color and over-plotted horizontal velocity vectors at a height of 1 $\mathrm{Mm}$ above the mean solar surface.
Blue and yellow squares display the regions displayed in the left and right column, respectively, of Fig. \ref{fig:my_label2}.

To quantify the energy contribution by vortices, we first calculate total vertical Poynting flux in the selected vortex regions at full resolution as well as for the degraded resolutions.
Total vertical Poynting flux at the selected vortex regions is displayed in Fig. \ref{fig:my_label6}(a).
We also calculate the fractional Poynting flux over vortices by taking the ratio of total Poynting flux over all the vortices to the total Poynting flux over the whole computational box at the same geometrical height as shown in Fig. \ref{fig:my_label6}(b)\footnote{Note that the fractional Poynting flux in vortices can be greater than unity because the signed Poynting flux is used, which mainly points upwards in vortices, but is more evenly balanced outside them.}.
For comparison, we have also displayed the area coverage of selected vortex locations in red.
We find that the Poynting flux contribution of vortices is much higher than their area coverage at full resolution simulations as well as for the degraded resolutions particularly in the lower chromosphere.
Moreover, to confirm that this excess Poynting flux is due to rotational motion (and not, e.g., to the fact that vortices are mainly located in magnetic concentrations), we calculate the mean magnetic field over selected vortex regions (dashed curve) and compare it with the total horizontally averaged magnetic field (solid curve) at full resolution simulations as shown in Fig. \ref{fig:my_label6}(c).
Since there is no excess magnetic field concentrated over vortices above $\sim$ 700 $\mathrm{km}$, the additional energy transport at vortex locations can essentially be attributed to the rotational motions therein. 

To analyze the vortex contribution to the energy requirement at various layers of the solar atmosphere, we calculate the Poynting flux density i.e. $\frac{\int_{a}S_z \,ds}{\int_A \,ds}$. Here `a' refers to the area covered by vortices while `A' corresponds to the horizontal area of the computational domain.
At full resolution, the Poynting flux density due to vortices is 33.5 $\mathrm{kWm^{-2}}$, 12.5 $\mathrm{kWm^{-2}}$ and 7.5 $\mathrm{kWm^{-2}}$ at 500 $\mathrm{km}$, 1000 $\mathrm{km}$ and 1500 $\mathrm{km}$, respectively above the mean solar surface while the estimated energy flux requirement for plage in the lower, middle and upper chromosphere is $>10$ $\mathrm{kWm^{-2}}$, 10 $\mathrm{kWm^{-2}}$ and 2 $\mathrm{kWm^{-2}}$ respectively (\citealt{withbroe1977}).
\section{Discussion and conclusion}\label{4}
Vortex flows exist over various spatial and temporal scales throughout the solar atmosphere and are of great importance due to their potential in twisting the magnetic field lines and hence facilitating Poynting flux transport (\citealt{Shelyag_2012,Wedemeyer-Bohm2012}).
It is crucial to investigate small-scale vortices since they are more abundant and rotate faster than the larger vortices (\citealt{Kato2017,Liu_2019,Giagkiozis_2018}).
We performed high-resolution simulations of a unipolar solar plage region to detect vortices at various scales and investigate their importance in the net energy transport into and through the chromosphere at different spatial scales.
We used the swirling strength criterion for vortex identification and detected vortices with diameters as small as 50-100 $\mathrm{km}$ at the solar surface.
Such small-scale vortices have not yet been identified in observations due to insufficient spatial resolution and unreliable identification methods.

Most of the previous observational and simulation studies of vortex flows pertain to quiet Sun conditions.
Using CO$^5$BOLD simulation code \citet{Wedemeyer-Bohm2012} detected vortex flows (through visual inspection) of similar widths as found in observations.
\citet{Kato2017} applied an automated detection technique to the simulation data as used by \citet{Wedemeyer-Bohm2012} and reported that the majority of the detected vortices are much smaller than found in observations.
They considered the smaller vortices, coinciding with a comparatively larger vortex flow, as artificial detections and attributed them to the larger vortex flow.
In the present study, we show that both small-scale and larger vortex flows are co-existing and vortex detection methods (based on velocity gradients) may still identify larger vortex flows if combined with smearing of data.
Since the smaller vortices are in fact the main contributor to the Poynting flux, they should not be neglected.

It was recently shown by \citet{kostas_2019} that a single vortex could have a fine structure and may consist of multiple smaller vortices, a validation through simulation was missing. 
The present work addresses the question of scaling of vortex flows ranging from a few tens of kilometers to Megameters.
We demonstrated that a large vortex, as seen at comparatively low spatial resolution, consists of a large number of smaller vortices, when seen at high spatial resolution.

We compared the contribution of the horizontal and the vertical plasma motions to the total vertical Poynting flux.
We found horizontal plasma motions as the main contributor, which implicitly indicates the importance of vortices in energy transport.
Our simulation setup is similar to that of \citet{shelyag_2011} and \citet{Shelyag_2012}, and our results related to Poynting flux are in agreement with their findings.
Moreover, they demonstrated that vortex flows primarily coincide with regions of strong magnetic fields.
This is true in our case as well in the photosphere, however, there is no excess concentration of the magnetic field at the vortex locations in the chromosphere. 
In our simulations, we find localized magnetic field concentrations in the photosphere and lower chromosphere.
The field in these concentrations expands with height and they 
merge at a height between 500-700 km above the mean surface.
Above this merging height, the magnetic field becomes increasingly homogeneous.
This is consistent with the observations of plage regions (\citealt{Buente_1993}).
Therefore, in the higher atmospheric layers, large vertical Poynting flux over vortices can be attributed to the horizontal plasma motions associated with them.
We further investigated the scaling behaviour of the total vertical Poynting flux as a function of effective spatial resolution and found it to follow an exponential power law.
It shows that there can be a ``hidden'' Poynting flux associated with small-scale flows, in particular with vortices, which is not currently resolved by observations.
This invisible energy flux can be estimated with the scaling law.
At full resolution, the Poynting flux density due to vortices over the area of the computational box (12 $\mathrm{Mm}$ $\times$ 12 $\mathrm{Mm}$) is higher than the estimated energy flux required to heat the chromosphere in active regions.
Thus we demonstrate that vortex flows have the potential to heat the chromosphere of a plage.
\section{Acknowledgements}
The authors acknowledge L. P. Chitta and D. Przybylski for useful discussions on various aspects of this paper. This project has received funding from the European Research Council (ERC) under the European Union’s Horizon 2020 research and innovation programme (grant agreement No. 695075) and has been supported by the BK21 plus program through the National Research Foundation (NRF) funded by the Ministry of Education of Korea. N. Y. thanks ISSI Bern for support for the team "The Nature and Physics of Vortex Flows in Solar Plasmas".



\bibliography{paper}

\begin{thebibliography}{}
\expandafter\ifx\csname natexlab\endcsname\relax\def\natexlab#1{#1}\fi
\providecommand{\url}[1]{\href{#1}{#1}}
\providecommand{\dodoi}[1]{doi:~\href{http://doi.org/#1}{\nolinkurl{#1}}}
\providecommand{\doeprint}[1]{\href{http://ascl.net/#1}{\nolinkurl{http://ascl.net/#1}}}
\providecommand{\doarXiv}[1]{\href{https://arxiv.org/abs/#1}{\nolinkurl{https://arxiv.org/abs/#1}}}

\bibitem[{{Abbett} \& {Fisher}(2012)}]{abbett_2012}
{Abbett}, W.~P., \& {Fisher}, G.~H. 2012, \solphys, 277, 3,
  \dodoi{10.1007/s11207-011-9817-3}

\bibitem[{{Attie} {et~al.}(2009){Attie}, {Innes}, \& {Potts}}]{Attie2009}
{Attie}, R., {Innes}, D.~E., \& {Potts}, H.~E. 2009, \aap, 493, L13,
  \dodoi{10.1051/0004-6361:200811258}

\bibitem[{{Bonet} {et~al.}(2008){Bonet}, {M{\'a}rquez}, {S{\'a}nchez Almeida},
  {Cabello}, \& {Domingo}}]{Bonet2008}
{Bonet}, J.~A., {M{\'a}rquez}, I., {S{\'a}nchez Almeida}, J., {Cabello}, I., \&
  {Domingo}, V. 2008, The Astrophysical Journal Letters, 687, L131,
  \dodoi{10.1086/593329}

\bibitem[{{Bonet} {et~al.}(2010){Bonet}, {M{\'a}rquez}, {S{\'a}nchez Almeida},
  {Palacios}, {Mart{\'\i}nez Pillet}, {Solanki}, {del Toro Iniesta}, {Domingo},
  {Berkefeld}, {Schmidt}, {Gandorfer}, {Barthol}, \& {Kn{\"o}lker}}]{bonet2010}
{Bonet}, J.~A., {M{\'a}rquez}, I., {S{\'a}nchez Almeida}, J., {et~al.} 2010,
  The Astrophysical Journal, 723, L139, \dodoi{10.1088/2041-8205/723/2/L139}

\bibitem[{{Brandt} {et~al.}(1988){Brandt}, {Scharmer}, {Ferguson}, {Shine},
  {Tarbell}, \& {Title}}]{1988brandt}
{Brandt}, P.~N., {Scharmer}, G.~B., {Ferguson}, S., {et~al.} 1988, \nat, 335,
  238, \dodoi{10.1038/335238a0}

\bibitem[{{B\"unte} {et~al.}(1993){B\"unte}, {Solanki}, \&
  {Steiner}}]{Buente_1993}
{B\"unte}, M., {Solanki}, S.~K., \& {Steiner}, O. 1993, \aap, 268, 736

\bibitem[{de~Souza~e Almeida~Silva {et~al.}(2018)de~Souza~e Almeida~Silva,
  Rempel, Gomes, Requerey, \& Chian}]{Silva_2018}
de~Souza~e Almeida~Silva, S., Rempel, E.~L., Gomes, T. F.~P., Requerey, I.~S.,
  \& Chian, A. C.-L. 2018, The Astrophysical Journal, 863, L2,
  \dodoi{10.3847/2041-8213/aad180}

\bibitem[{{Evershed}(1909)}]{1909evershed}
{Evershed}, J. 1909, \mnras, 69, 454, \dodoi{10.1093/mnras/69.5.454}

\bibitem[{Fedun {et~al.}(2011)Fedun, Shelyag, Verth, Mathioudakis, \&
  Erd\'elyi}]{Fedun2011}
Fedun, V., Shelyag, S., Verth, G., Mathioudakis, M., \& Erd\'elyi, R. 2011,
  Annales Geophysicae, 29, 1029, \dodoi{10.5194/angeo-29-1029-2011}

\bibitem[{Giagkiozis {et~al.}(2018)Giagkiozis, Fedun, Scullion, Jess, \&
  Verth}]{Giagkiozis_2018}
Giagkiozis, I., Fedun, V., Scullion, E., Jess, D.~B., \& Verth, G. 2018, The
  Astrophysical Journal, 869, 169, \dodoi{10.3847/1538-4357/aaf797}

\bibitem[{{Kato} \& {Wedemeyer}(2017)}]{Kato2017}
{Kato}, Y., \& {Wedemeyer}, S. 2017, \aap, 601, A135,
  \dodoi{10.1051/0004-6361/201630082}

\bibitem[{{Khomenko} \& {Collados}(2012)}]{khomenko_2012}
{Khomenko}, E., \& {Collados}, M. 2012, \apj, 747, 87,
  \dodoi{10.1088/0004-637X/747/2/87}

\bibitem[{{Khomenko} {et~al.}(2018){Khomenko}, {Vitas}, {Collados}, \& {de
  Vicente}}]{khomenko_2018}
{Khomenko}, E., {Vitas}, N., {Collados}, M., \& {de Vicente}, A. 2018, \aap,
  618, A87, \dodoi{10.1051/0004-6361/201833048}

\bibitem[{Kitiashvili {et~al.}(2012)Kitiashvili, Kosovichev, Mansour, \&
  Wray}]{Kitiashvili_2012}
Kitiashvili, I.~N., Kosovichev, A.~G., Mansour, N.~N., \& Wray, A.~A. 2012, The
  Astrophysical Journal, 751, L21, \dodoi{10.1088/2041-8205/751/1/l21}

\bibitem[{Liu {et~al.}(2019)Liu, Nelson, \& Erd{\'{e}}lyi}]{Liu_2019}
Liu, J., Nelson, C.~J., \& Erd{\'{e}}lyi, R. 2019, The Astrophysical Journal,
  872, 22, \dodoi{10.3847/1538-4357/aabd34}

\bibitem[{{Moll} {et~al.}(2012){Moll}, {Cameron}, \&
  {Sch{\"u}ssler}}]{moll2012}
{Moll}, R., {Cameron}, R.~H., \& {Sch{\"u}ssler}, M. 2012, \aap, 541, A68,
  \dodoi{10.1051/0004-6361/201218866}

\bibitem[{Nordlund(1985)}]{Nordlund1985}
Nordlund, {\AA}. 1985, Solar Physics, 100, 209, \dodoi{10.1007/BF00158429}

\bibitem[{Park {et~al.}(2016)Park, Tsiropoula, Kontogiannis, Tziotziou,
  Scullion, \& Doyle}]{ParkS2016}
Park, S, H., Tsiropoula, G., Kontogiannis, I., {et~al.} 2016, A{\&}A, 586, A25,
  \dodoi{10.1051/0004-6361/201527440}

\bibitem[{Rempel(2014)}]{Rempel2014}
Rempel, M. 2014, The Astrophysical Journal, 789, 132,
  \dodoi{10.1088/0004-637x/789/2/132}

\bibitem[{{Rempel}(2017)}]{Rempel2017}
{Rempel}, M. 2017, \apj, 834, 10, \dodoi{10.3847/1538-4357/834/1/10}

\bibitem[{Requerey {et~al.}(2018)Requerey, Cobo, Gošić, \&
  Bellot~Rubio}]{iker2018}
Requerey, I.~S., Cobo, B.~R., Gošić, M., \& Bellot~Rubio, L.~R. 2018, A{\&}A,
  610, A84, \dodoi{10.1051/0004-6361/201731842}

\bibitem[{Shelyag {et~al.}(2011)Shelyag, Fedun, Keenan, Erd\'elyi, \&
  Mathioudakis}]{shelyag_2011}
Shelyag, S., Fedun, V., Keenan, F.~P., Erd\'elyi, R., \& Mathioudakis, M. 2011,
  Annales Geophysicae, 29, 883, \dodoi{10.5194/angeo-29-883-2011}

\bibitem[{Shelyag {et~al.}(2016)Shelyag, Khomenko, de~Vicente, \&
  Przybylski}]{Shelyag_2016}
Shelyag, S., Khomenko, E., de~Vicente, A., \& Przybylski, D. 2016, The
  Astrophysical Journal, 819, L11, \dodoi{10.3847/2041-8205/819/1/l11}

\bibitem[{Shelyag {et~al.}(2012)Shelyag, Mathioudakis, \&
  Keenan}]{Shelyag_2012}
Shelyag, S., Mathioudakis, M., \& Keenan, F.~P. 2012, The Astrophysical
  Journal, 753, L22, \dodoi{10.1088/2041-8205/753/1/l22}

\bibitem[{{Steiner} {et~al.}(2008){Steiner}, {Rezaei}, {Schaffenberger}, \&
  {Wedemeyer-B{\"o}hm}}]{steiner_2008}
{Steiner}, O., {Rezaei}, R., {Schaffenberger}, W., \& {Wedemeyer-B{\"o}hm}, S.
  2008, \apjl, 680, L85, \dodoi{10.1086/589740}

\bibitem[{{Su} {et~al.}(2012){Su}, {Wang}, {Veronig}, {Temmer}, \&
  {Gan}}]{Su2012}
{Su}, Y., {Wang}, T., {Veronig}, A., {Temmer}, M., \& {Gan}, W. 2012, The
  Astrophysical Journal Letters, 756, L41, \dodoi{10.1088/2041-8205/756/2/L41}

\bibitem[{{Tziotziou} {et~al.}(2019){Tziotziou}, {Tsiropoula}, \&
  {Kontogiannis}}]{kostas_2019}
{Tziotziou}, K., {Tsiropoula}, G., \& {Kontogiannis}, I. 2019, \aap, 623, A160,
  \dodoi{10.1051/0004-6361/201834679}

\bibitem[{Vögler {et~al.}(2005)Vögler, Shelyag, Schüssler, Cattaneo, Emonet,
  \& Linde}]{voegler2005}
Vögler, A., Shelyag, S., Schüssler, M., {et~al.} 2005, A{\&}A, 429, 335,
  \dodoi{10.1051/0004-6361:20041507}

\bibitem[{{Wedemeyer} {et~al.}(2013){Wedemeyer}, {Scullion}, {Rouppe van der
  Voort}, {Bosnjak}, \& {Antolin}}]{Wedemeyer2013}
{Wedemeyer}, S., {Scullion}, E., {Rouppe van der Voort}, L., {Bosnjak}, A., \&
  {Antolin}, P. 2013, \apj, 774, 123, \dodoi{10.1088/0004-637X/774/2/123}

\bibitem[{Wedemeyer-B{\"{o}}hm {et~al.}(2012)Wedemeyer-B{\"{o}}hm, Scullion,
  Steiner, {Van Der Voort}, {De La Cruz Rodriguez}, Fedun, \&
  Erd{\'{e}}lyi}]{Wedemeyer-Bohm2012}
Wedemeyer-B{\"{o}}hm, S., Scullion, E., Steiner, O., {et~al.} 2012, Nature,
  486, 505, \dodoi{10.1038/nature11202}

\bibitem[{{Withbroe} \& {Noyes}(1977)}]{withbroe1977}
{Withbroe}, G.~L., \& {Noyes}, R.~W. 1977, Annual Review of Astronomy and
  Astrophysics, 15, 363, \dodoi{10.1146/annurev.aa.15.090177.002051}

\bibitem[{{Yan} {et~al.}(2008){Yan}, {Qu}, \& {Xu}}]{2008yan}
{Yan}, X.~L., {Qu}, Z.~Q., \& {Xu}, C.~L. 2008, The Astrophysical Journal, 682,
  L65, \dodoi{10.1086/590953}

\bibitem[{Zhou {et~al.}(1999)Zhou, Adrian, Balachandar, \& Kendall}]{zhou1999}
Zhou, J., Adrian, R.~J., Balachandar, S., \& Kendall, T.~M. 1999, Journal of
  Fluid Mechanics, 387, 353, \dodoi{10.1017/S002211209900467X}

\end{thebibliography}
\bibliographystyle{aasjournal}



\end{document}